\documentclass[12pt,a4paper]{article} 
\usepackage[dvipsnames,usenames]{color}
\usepackage{graphicx}
\usepackage[normalem]{ulem}
\usepackage{lscape}
\usepackage{multirow}
\usepackage{ftnxtra}

\begin{document}

\title{Reply to: ``TOV or OV? The Whole Story''}

\author{\.{I}brahim 
Semiz\thanks{mail: ibrahim.semiz@boun.edu.tr} \\ \\
Bo\u{g}azi\c{c}i University, Department of Physics\\
Bebek, \.{I}stanbul, TURKEY}
    
\date{ }

\maketitle

\begin{abstract}
We argued previously that the well-known equation for hydrostatic equilibrium in a static spherically symmetric spacetime supported by an isotropic perfect fluid should be called the Oppenheimer-Volkoff (OV) equation, rather than the Tolman-Oppenheimer-Volkoff (TOV) equation; a later ArXiv manuscript has disagreed. Here we reply to that comment, reaffirming the  original argument.
\end{abstract}

\vspace{10 mm}
In previous work~\cite{OvTovSemiz}, by scrutinizing the usually cited literature~\cite{Tolman,OV} we argued that the well-known equation for hydrostatic equilibrium in a static spherically symmetric spacetime supported by an isotropic perfect should be called the Oppenheimer-Volkoff (OV) equation, rather than the Tolman-Oppenheimer-Volkoff (TOV) equation. In a later {\it ArXiv} manuscript~\cite{TovOvHerrera}, Herrera argued that the (T)OV equation is a ``rather elementary'' extension of the equation of local conservation of energy-momentum (LCEM), which appears in two 1930 works of Tolman~\cite{Tolman30a,Tolman30b}, and also in the famous 1933 paper~\cite{Lemaitre} of Lema\^{i}tre (where anisotropic fluids are also included); hence even more credit is due Tolman, possibly also due Lema\^{i}tre, in addition to Oppenheimer \& Volkoff.

  Let us recall that we are interested in  finding valid solutions for the contents and structure of a static spherically symmetric  spacetime filled with an isotropic perfect fluid (SSSPF solutions), that satisfy Einstein's Field Equations (EFE). The relevant conventions and equations were given in~\cite{OvTovSemiz}; but here we repeat the expression for the line element, since the particular form defines the Schwarzschild coordinate system, which will be compared with the alternative system used in Tolman's 1930 works~\cite{Tolman30a,Tolman30b}: 
\begin{equation}
ds^{2} = -B(r) dt^{2} + A(r) dr^{2} + r^{2} d\Omega^{2}.   \label{ansatz}
\end{equation}
Here the metric functions $B(r)$ and $A(r)$ are two of the four fuctions comprising the solution, $\rho(r)$ and $p(r)$ being the other two.

In what follows, we explain what convenience the OV equation brings on top of using the   LCEM equation, in conjunction with the Equation of State (EoS) and the components of the EFE. We also scrutinize the references quoted by Herrera, and argue that their content is not suitable for attempting exact solutions in this context, and they do not mount any such attempt. Hence we argue that the criticism is not justified, and our original argument stands.


\section{SSSPF Solutions without the OV Formalism}
\label{SSSPFnoOV}

In principle, the problem can be solved by
\begin{eqnarray}
A' & = & \frac{A^{2}}{r} \left(\kappa \rho r^{2} - 1 + \frac{1}{A} \right),\label{EFE00-S} \\ 
B'  & = & \frac{B}{r} \left(\kappa A p r^{2} - 1 + A \right), \label{EFE11-S} \\ 
p' & = & -2\frac{B'}{B}(\rho+p), \label{LCEM} \\
f(p,\rho) & = & 0   \;\;\;\;\;\;\;\;\;\;\;\; {\rm (``EoS'')} \label{EoS} 
\end{eqnarray}
where (\ref{EFE00-S}), (\ref{EFE11-S}) and (\ref{LCEM}) are taken from (6),  (7) and (9) of \cite{OvTovSemiz}, respectively. We used the simpler LCEM equation instead of the complicated 22-component of the EFE (also because EFE-22 is second-order).

These  constitute a closed system of \underline{four} equations, with \underline{three} {\it first order} differential equations; which {\it could} in principle be integrated starting from the origin, for example. However, while the values of $\rho$ and $p$ at the origin have easy physical interpretation as the central density and pressure, the central values of $A$ and $B$ are not obvious even for a regular solution (however, see Sect.\ref{sect:Conc}).


\section{SSSPF Solutions with the OV Formalism}
\label{sect:Tolman}

  The OV formalism introduces a new function $F$ proportional to the mass function, so that five equations are needed. They are, from Sect.4 of~\cite{OvTovSemiz}:
\begin{eqnarray}
F' & = & \kappa  \rho r^{2}, 		\label{F_def} \\ 
p' & = & - \frac{(\kappa p r^{3} + F)}{2 r (r-F)} (\rho + p).     \label{OV} \\
f(p,\rho) & = & 0   \;\;\;\;\;\;\;\;\;\;\;\; {\rm (``EoS'')} \nonumber \\
A  & = & \frac{r}{r-F}, \label{A-def} \\ 
\frac{B'}{B} & = & \frac{\kappa p r^{2} + 1}{r-F} - \frac{1}{r},       \label{B'/B}
\end{eqnarray}

At first sight, it would seem that introducing a new function and a new equation would make the problem more complicated. However, now the first \underline{three} equations constitute a closed system, with \underline{two} {\it first order} differential equations; so the system needed to be immediately integrated is simpler. Furthermore, the value of $F$  at the origin is obvious, zero, for  a regular solution, in addition to $\rho(0)$ and $p(0)$ being the central density and pressure, respectively. In fact, in this formalism, it can be seen that $F$ cannot diverge at the origin, if one requires that the mass there not be infinite, even if the density does (for solutions where $A$ and $B$ stay positive for all values of $r$).


\section{Conclusions}
\label{sect:Conc}

In the previous paper~\cite{OvTovSemiz}, we argued that the work~\cite{Tolman} does not justify prepending a T to OV. The Herrera comment~\cite{TovOvHerrera} agrees with the judgement on ~\cite{Tolman}, but argues that the essence of the (T)OV formalism lies in the LCEM equation (\ref{LCEM}) which appeared in the literature in earlier\footnote{It is possible that further search could uncover this equation in even earlier work.} Tolman papers~\cite{Tolman30a,Tolman30b}. We would like to also remark at this point that none of the early work in this context cites these, including the author himself~\cite{Tolman}; when we scrutinized that early work for \cite{OvTovSemiz}, we found that usually \cite{Tolman} is cited. This is also true for the part of the recent work that we have seen.

We point out that without introducing the function $F$ \`{a} la Oppenheimer \& Volkoff, the system to be solved consists of three first-order differential and one algebraic equations, with the boundary condition (bc) of at least one variable not obvious. With the introduction of $F$, the problem immediately to be solved becomes simpler: two first-order differential and one algebraic equations, with the bc's condition of all three variables obvious (Actually, after the introduction of $F$, the bc for $A$ becomes clear: $F(r) \approx (4/3) \pi \rho_{0} r^{3}$
for small $r$, so (\ref{A-def}) gives $A(0)=1$; but this becomes clear {\it only} after the introduction of $F$).

Furthermore, none of \cite{Tolman30a}, \cite{Tolman30b} or \cite{Lemaitre} attempt to reach an analytical solution\footnote{this despite the fact that \cite{Lemaitre} recognizes a mass funtion.}, or point the way to one, or even describe how a numerical solution may be attained. In fact, in the isotropic coordinates used in \cite{Tolman30a,Tolman30b}, where the line element is equivalent to
\begin{equation}
ds^{2} = -B(r) dt^{2} + A(r) \left[dr^{2} + r^{2} d\Omega^{2}\right],   \label{isotrpic_ansatz}
\end{equation}
two of the three components of the EFE are second-order in the metric variables, so even if one of them is replaced by the LCEM equation, there is still at least one second-order equation left, making making analytical or numerical solution difficult.

Hence, we conclude that both the introduction of the ($\sim$mass) function $F$, and the use of Schwarzschild coordinates, by Oppenheimer \& Volkoff represent decicive steps in formulating the algorithm towards solutions of the problem at hand, so we stand by our original conclusion, that the relevant equation be called  ``The Oppenheimer-Volkoff (OV) equation''.

\end{document}